\documentclass[a4paper,11pt]{article}
\pdfoutput=1 

\usepackage{jcappub} 
\usepackage[x11names]{xcolor}
\usepackage{colortbl,hhline}
\usepackage[T1]{fontenc} 

\definecolor{phthaloblue}{rgb}{0.0, 0.06, 0.54}

\definecolor{Green}{rgb}{0,100,0}
\definecolor{Blue}{rgb}{0,0,100}
\definecolor{somered}{rgb}{0.7, 0.06, 0.1}

\title{\boldmath Distinguishing cosmologies using the turn-around radius near galaxy clusters}

\author[a]{Steen H. Hansen}
\author[b]{Farbod Hassani,}
\author[b]{Lucas Lombriser,}
\author[b]{Martin Kunz}

\affiliation[a]{DARK, Niels Bohr Institute, University of Copenhagen, Lyngbyvej 2, 2100 Copenhagen {\O}, Denmark}
\affiliation[b]{D\'epartement de Physique Th\'eorique, Universit\'e de Gen\`eve, 24 quai Ernest Ansermet, 1211 Gen\`eve 4, Switzerland}

\emailAdd{hansen@nbi.ku.dk}
\emailAdd{Farbod.Hassani@unige.ch}
\emailAdd{Lucas.Lombriser@unige.ch}
\emailAdd{martin.kunz@unige.ch}

\abstract{Outside galaxy clusters the competition between
the inwards gravitational attraction and the outwards
expansion of the Universe leads to a special radius of
velocity cancellation, which is called the turn-around radius. Measurements
of the turn-around radius hold promises of constraining
cosmological parameters, and possibly even properties of
gravity. Such a measurement is, however, complicated by the
fact that the surroundings of galaxy clusters are not spherical,
but instead are a complicated collection of filaments, sheets
and voids. 
In this paper we use the results of numerically
simulated universes to quantify realistic error-bars of the
measurement of the turn-around radius. We find that for 
a $\Lambda$CDM cosmology these error-bars are typically of
the order of $20\%$.
We numerically simulate three different implementations of dark energy models and of a scalar dark sector interaction
to address whether the turn-around radius can be used to constrain
non-trivial cosmologies, and we find that only rather extreme
models can be distinguished from a $\Lambda$CDM universe due to
the large error-bars arising from the non-trivial cluster environments.}

\begin{document}
\maketitle
\flushbottom

\section{Introduction}
\label{sec:intro}

The turn-around radius is the unique distance where the gravitational pull
of large cosmological structures exactly cancels the expansion of the 
Universe. It therefore provides a special place to constrain
the 
properties of the expanding universe, for instance the amount of dark
energy \citep{2014JCAP...05..017P,2014JCAP...09..020P}, or
the force of gravity from the gravitational structures
\citep{2016PDU....11...11F,2018IJMPD..2748006C}.
The main observational difficulty with a measurement of the
turn-around radius is that the 3-dimensional position of galaxies
is very difficult to obtain, since only the 2-dimensional position
on the sky is readily observable. 
{For very nearby objects it may be possible to measure the
turn-around radius directly \citep{hoffman2018}.}
This complication {at cosmological distances} may be
overcome if one could measure a coherent motion of some of the galaxies. One 
such possibility was suggested in \citep{2014MNRAS.442.1887F},
where it was demonstrated that galaxies in large 2-dimensional sheets
in their early phase of gravitational collapse indeed have
properties allowing one to determine the full 3-dimensional 
spatial distribution. From numerical
simulations it is known that these large structures are Zeldovich
pancakes (also called sheets), which are over-densities that have only collapsed
along the one dimension \citep{2015MNRAS.447.1333W, 2016JCAP...04..007B}.
In reference \citep{2015ApJ...815...43L} it was proposed that it is possible to use a detection of such a sheet to actually measure the
turn-around radius. This method has subsequently been investigated 
in a series of papers, in order to measure properties either of 
the clusters or of the expanding universe
\cite{2015ApJ...807..122L, 2016ApJ...832..185L, 2018ApJ...856...57L, 2016ApJ...832..123L, 2017ApJ...839...29L, 2017ApJ...842....2L, 2016MNRAS.455.2267R}.
{It was recently suggested that $\Lambda$CDM model and an f(R) model
of modified gravity could fairly easily be distinguished in the
future, by measuring the turn-around radius and the virial mass \citep{Lopes:2018jcw} (see
also \citep{Lopes:2018uhq,Capozziello:2018oiw}).}
Most of the analyses mentioned above assume that the gravitational potential
of the large cosmological structures are approximately spherical, and that the
measured turn-around radius in a given direction therefore provides a fair representation
of the turn-around radius of the galaxy cluster. 
{The departure from sphericity around a galaxy cluster does, however,
induce a large scatter in the measured turn-around radius. This is
because the coherently moving galaxies (Zeldovich pancakes) which are
used to measure the turn-around radius, are quite localized in space,
and hence highly directional.}
In this paper we will first of all check to which degree this is an accurate
approach, and at the same time we will quantify the magnitude of the error-bar
of the measured turn-around radius. It turns out that the corresponding 
error-bars are significant, and must be included in future analyses.
We then use this result to evaluate to which degree one can actually use
measurements of turn-around radii to constrain alternative cosmologies. 
As concrete examples we consider the numerical implementations of three dark energy models and
of a scalar dark sector interaction, which can be compared with the
standard $\Lambda$CDM cosmology. We
demonstrate that a correct inclusion of the systematic error-bars
is very important, and makes it rather difficult to distinguish between
different cosmologies.


\section{Turn-around radius}

\begin{figure}[ptb]
\centering 
\includegraphics[width=.45\textwidth,origin=c,angle=0]{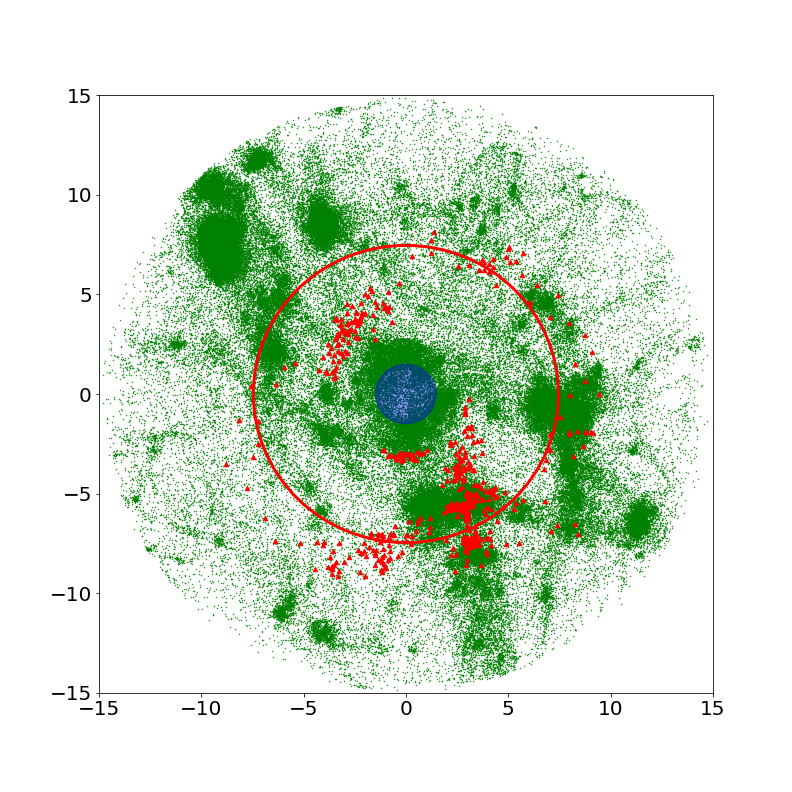}
\hfill
\includegraphics[width=.45\textwidth,origin=c,angle=0]{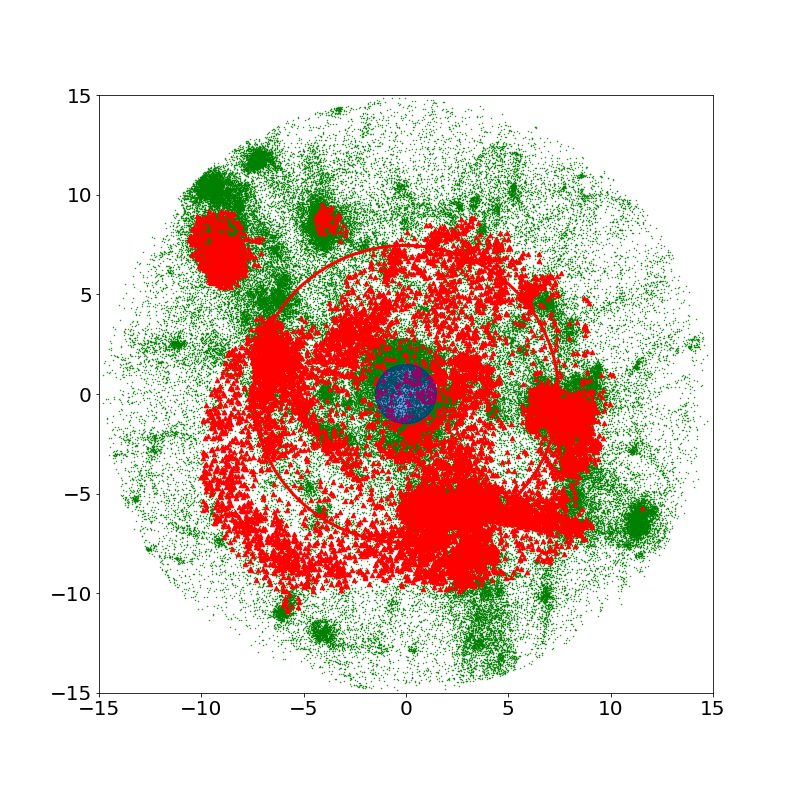}
\caption{\label{fig:rta} This figure exemplifies that the 
turn-around radius is hard to identify uniquely. 
The green dots represent all particles out to 
10 virial radii around a large galaxy cluster. The central blue region is
one virial radius, and the red circle is a guide-the-eye line. The red, triangular symbols
represent galaxies which happen to have zero radial velocity with respect to
the central galaxy cluster. The corresponding radius is the turn-around radius.
Along directions with massive substructures the
potential is highly non-trivial (and non-spherical) and hence the turn-around
radius depends on the direction in which it is measured. Left panel: 
The  zero radial velocity
galaxies (colour-coded red) 
are selected from a thin slice perpendicular to the line-of-sight.
Right panel: All the (almost spherically distributed)
galaxies with zero radial velocity are colour-coded red.}
\end{figure}

Figure~\ref{fig:rta} exemplifies the non-triviality of uniquely
defining the turn-around radius for realistic cosmological structures.
The green dots show particles within 10 virial radii of the cluster.
The larger, red triangles show particles which have the property that
they have zero radial velocity with respect to the central cluster
(plus/minus 100 km/sec), and should thus represent particles at the
turn-around radius.
The red particles shown in the left panel are selected from a
thin slice of width half a virial radius, and we only select particles
outside of two virial radii, since the particles inside the virial
radius on average are all at rest with respect to the centre.  The
filled, central circle represents 1 virial radius.  The red circle is
a guide-the-eye line at 5 times the virial radii.

The problem is clearly seen on the left panel of figure~\ref{fig:rta}, 
namely that it is very difficult to
define a unique turn-around radius. First of all, in directions in
space with significant substructure, the turn-around radius may appear
significantly closer to the cluster, because the gravitational potential of the
large substructures affects the flow.

The second issue is, that it is not easy to observationally
select a spatial slice, because it is virtually impossible to measure
the line-of-sight distance to a given galaxy. Therefore a more realistic
representation would be the right panel in figure~\ref{fig:rta}. This figure
demonstrates the necessity of first identifying some coherence between
some of the galaxies, for instance by first finding a sheet, as proposed
in \citep{2014MNRAS.442.1887F}.

{The large substructures on the r.h.s. of Figure 1
  appear to have zero radial velocity. That is mainly an effect of the
  use of "large" symbols, and the fact that we here colour-code all
  particles with zero velocity plus/minus 100 km/sec.  This velocity
  range was chosen to make the infall galaxies visible on the
  l.h.s. of Figure.~1.  The substructures each have a significant
  internal velocity dispersion, and only a fraction of their particles
  happen to have zero radial velocity with respect to the direction
  towards the nearby galaxy cluster at this specific moment in time.}

{Some of the zero-radial-velocity particles may happen to be
splashback particles (returning towards the cluster after a recent
merger). Such particles are not likely to end up as coherently moving
galaxies (like in a Zeldovich pancake) so we have not studied this
further.}

\section{Finding the turn-around radius}

Measuring the turn-around radius requires a few steps~\cite{2015ApJ...815...43L}. 
The first is to find a collection of galaxies whose motion is somehow
correlated. The simplest choice is probably to select galaxies
which form part of a Zeldovich pancake
\citep{2014MNRAS.442.1887F}. These are
identified as lines in 
{observational
phase-space, which is the directly observationally space spanned by 
projected radial distance and line-of-sight velocity.}
Some of the
disadvantages of using the Zeldovich pancakes are that they are
almost invisible on the sky (their projected spatial 
over-density is quite low), and secondly that they must be
viewed at an angle between 20 and 70 degrees with respect to the
line-of-sight
\citep{2016JCAP...04..007B}. The advantage is, that once found,
the infall velocities of the galaxies belonging to the pancake is
coherent. This allows one to determine the viewing angle of the
pancakes, and thereby the actual radial distance to the nearby
galaxy cluster can be determined \cite{2014MNRAS.442.1887F}.

The next step is to use that the radial velocities of galaxies
near a galaxy cluster are the sum of two terms, namely the
expansion rate of the Universe, $v_H = r \, H$, and the 
peculiar velocity, $v_p$, which is negative due
to the attractive force of gravity
directed towards the
large nearby galaxy cluster. One thus has
\begin{equation}
v_r = v_H + v_p \, .
\label{eq:vrhp}
\end{equation}
It happens that the peculiar velocity typically follows the simple
form
\begin{equation}
v_p = - a \left( \frac{r_v}{r}\right)^b \, ,
\label{eq:vpab}
\end{equation}
where the coefficient $b$ is of the order $0.42$ for massive
galaxy clusters \cite{2014MNRAS.442.1887F}. This shape of the
peculiar velocity profile is often valid in the range
between 3 and 10 virial radii.
The virial radius, $r_v$, is here defined as $r_{200}$, namely
the radius within which the average density is 200 times the average 
density of the universe. The constant $a$ is a normalization to 
be determined.
The
detailed coefficients of equation~(\ref{eq:vpab}) are,
however, both dependent on the cluster mass and
redshift \cite{2014MNRAS.442.1887F, 2016ApJ...832..123L}.
In the analysis of this paper we will only use the fact
that the shape is given by equation~(\ref{eq:vpab}), and
we will even allow the coefficients to be different for
different directions around a given galaxy cluster.

The last step is now to solve eq.~(\ref{eq:vrhp}) for
$v_r=0$, which directly gives
us the turn-around radius \cite{2015ApJ...815...43L}.
In a future actual measurement one would also have to propagate the
error-bars on the Hubble parameter and on the measured galaxy
positions. For the present analysis we do not need to 
consider these.

{
In this paper we wish to measure the general scatter in the
turn-around radius near galaxy clusters, and we therefore use all 49
directions in space. This means that in this paper we do not need to
identify Zeldovich pancakes. Instead, we take the full region near
galaxy clusters directly from a numerical simulation, and solve
eqs.~(3.1) and (3.2) to find $v_r =0$ in 49 directions in space.}

\section{Spatial cones}
\begin{figure}[ptb]
\centering 
\includegraphics[width=.80\textwidth,origin=c,angle=0]{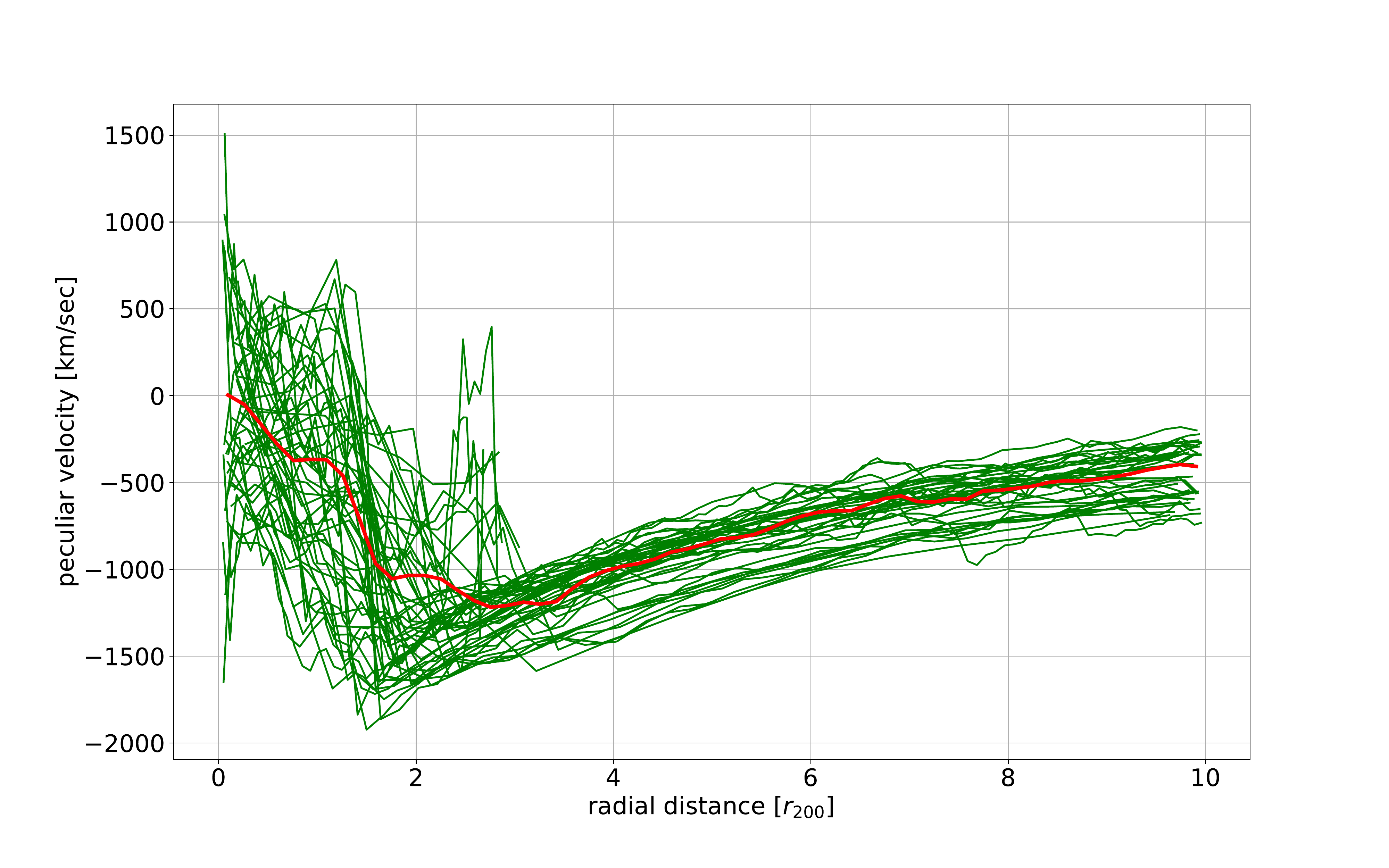}
\caption{\label{fig:pec1} Peculiar velocity as a function of radial distance. 
The 49 green lines each represent particles in a cone on the sky. The solid red
curve is the spherical average. This cluster is particularly well behaved and
equilibrated. 
{Within approximately 1 virial radius the average radial velocity is
zero, and therefore eq.~(3.1) tells us that the peculiar velocity on
average exactly cancels the Hubble expansion.}
Between approximately 1 and 3
virial radii there is infall towards the galaxy cluster (the total radial
velocity is negative), and toward larger
radii the peculiar velocity transitions slowly to zero.}
\end{figure}

For each numerically simulated galaxy cluster we wish to investigate the effect of
non-sphericity on the determined turn-around radius.
In order to quantify the variations along different directions in
space, we separate the sphere into $49$ cones of equal size. 
This fraction is chosen to resemble the fraction on the sky covered by a Zeldovich pancake 
\citep{2014MNRAS.442.1887F, 2016JCAP...04..007B}.
The
peculiar velocity of the particles in each cone are now averaged in
spherical bins, and the result is shown in figure \ref{fig:pec1}. The
solid, red curve is the spherical average of the full sphere. This 
figure shows a particularly
well-behaved and relaxed cluster, and therefore the infall profiles
are similar in all directions. In the innermost region (inside 1
or 2 virial radii) we see that the peculiar velocity equals minus the Hubble
expansion, in such a way that the average radial velocity is zero
in eq.~(\ref{eq:vrhp}).  At
large radii (from about 4 to 10 virial radii) the peculiar velocity is
seen to slowly go to zero. The spherical average is seen to represent
a fair average of the 49 cones.

\begin{figure}[tbp]
\centering 
\includegraphics[width=.80\textwidth,origin=c,angle=0]{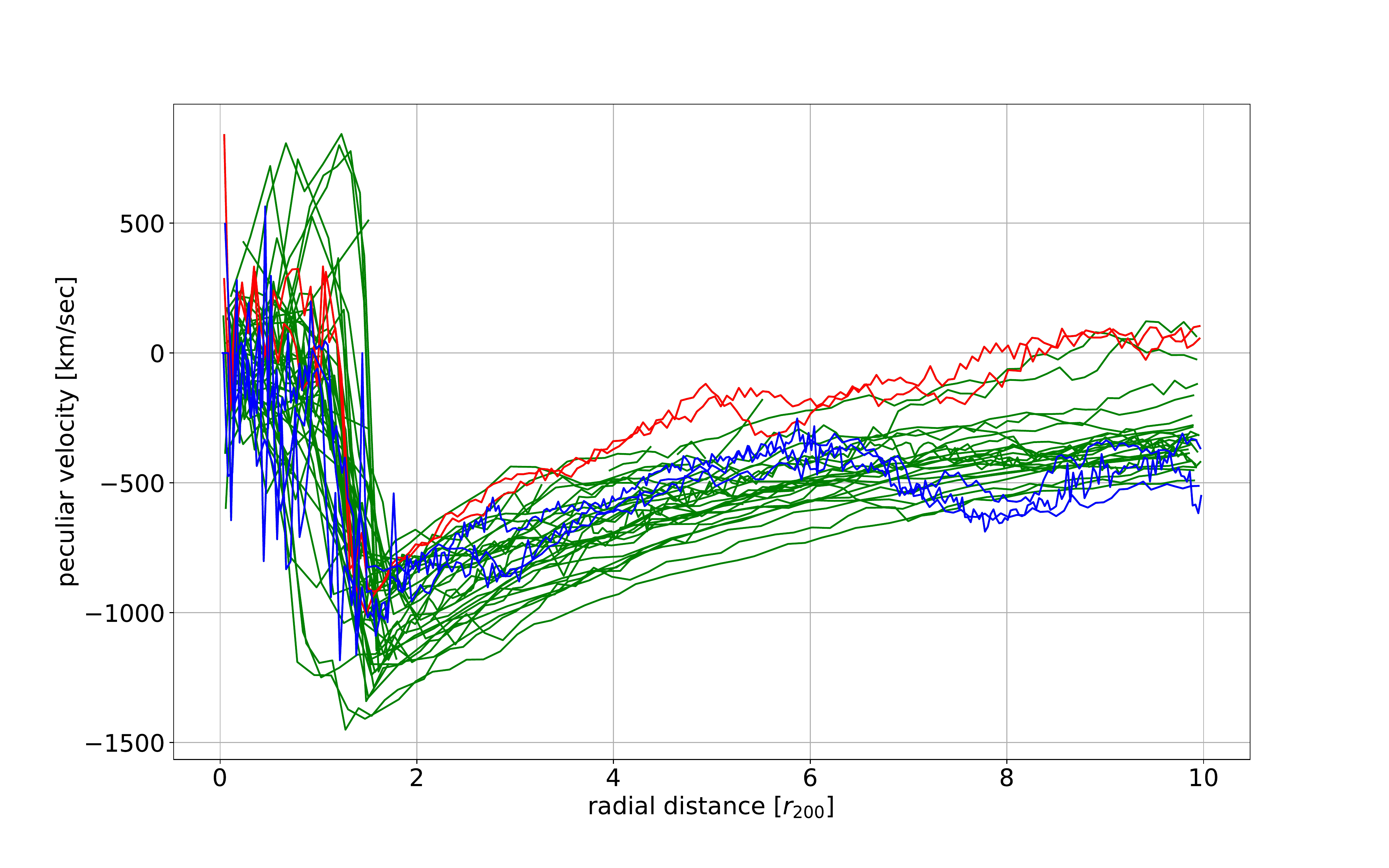}
\caption{\label{fig:pec2} Peculiar velocity as a function of radial distance. 
The 49 lines each represent particles in a cone on the sky. Many of the directions
are seen to behave similarly at large distances, however, a few directions stand
out: A few directions (red) even have positive peculiar velocities, and a few
have a clear transition (blue, here transitioning between 6 and 8 virial radii).
These non-trivial peculiar velocity profiles arise because of massive
sub-structures perturbing the overall potential.}
\end{figure}

\begin{figure}[tbp]
\centering 
\includegraphics[width=.45\textwidth,origin=c,angle=0]{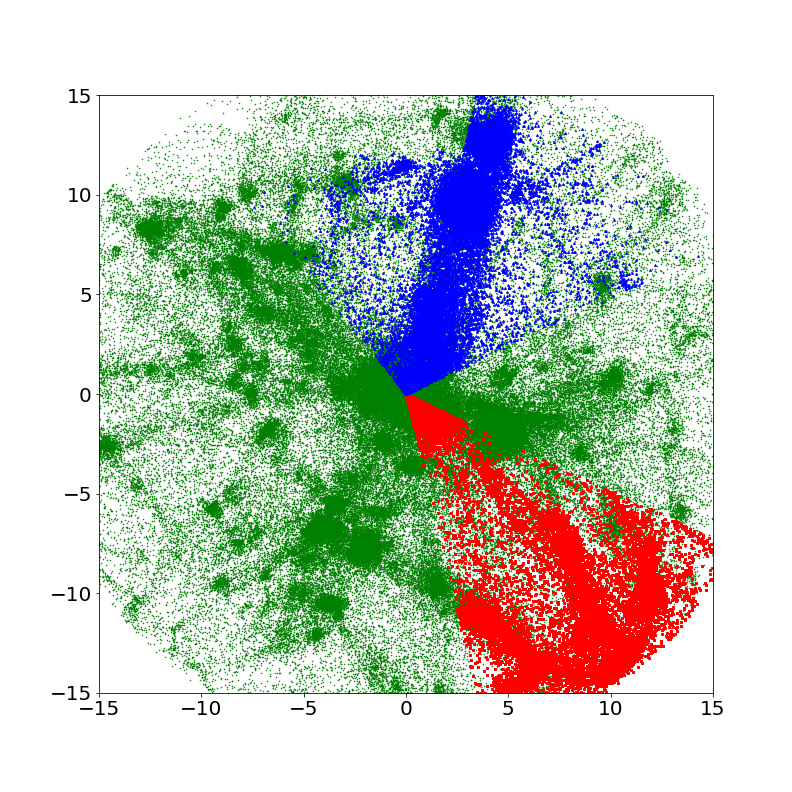}
\caption{\label{fig:directions} The spatial distribution of the particles belonging
to the peculiar velocity profiles of figure \ref{fig:pec2}. 
Some of the particles at 4 o'clock even have positive peculiar velocities.}
\end{figure}

A much more typical infall velocity picture is shown in figure
\ref{fig:pec2}. First of all we see a larger spread amongst the
velocity profiles of the 49 directions, and a few of the directions
are even seen to have {\em positive} peculiar velocities (crossing
zero around 8 virial radii for this specific cluster).  A few of these
directions are here color-coded red.  The cause of these positive
peculiar velocities are large nearby structures, whose potentials
significantly perturb the velocities of the particles in those
directions. This is clearly seen in figure \ref{fig:directions}, where
those regions are again color-coded red (triangles, at 4 o'clock).  
Another problematic peculiar
velocity curve is seen in figure \ref{fig:pec2} to depart from the
average trend around 6 virial radii, and becoming {\em larger} (more
negative) at increasing radii. A few of these directions are
color-coded blue. 
This is another feature of large, nearby substructures, as
clearly seen on figure \ref{fig:directions} (squares, at 1 o'clock).

When we identify Zeldovich pancakes on the sky, it is straightforward to avoid directions
in space which have large overdensities of galaxies. We therefore assume that these
directions have been identified, and we will remove them from the following analysis.
Concretely, for this paper we use a simplified approach, and
perform the following two tests. First, if the velocity
profile along a specific direction happens to have positive peculiar velocity, then
we remove that direction from the analysis (corresponding to the red lines
in figure \ref{fig:pec2}). Secondly, for any given direction
in space, we will fit a power-law to the peculiar velocity both between 
3-5 and 5-7 virial radii, and if any of the 
power-law coefficients are negative, then we remove that direction
(corresponding to the blue lines in figure \ref{fig:pec2}).

All the remaining directions will have slightly different infall profiles, and in
order to measure a realistic error-bar for the determined turn-around radius, we
will compare the variation amongst all these directions.

For each direction we now fit a power-law to the peculiar velocity in the radial
range $3-7$ virial radii 
\begin{equation}
v_p(r) =  - a \left( \frac{r_v}{r}\right) ^b \, .
\label{eq:powerlaw}
\end{equation}
Including slightly smaller/larger radii has very small effect on 
our conclusions. 
Since we know the full radial velocity is given by
\begin{equation}
v_r(r) =  r \, H - a \left( \frac{r_v}{r}\right) ^b \, ,
\label{eq:radialvelocity}
\end{equation}
we can find the turn-around radius by solving $v_r(r) =0$ for the radius \cite{2015ApJ...815...43L}.
For each cluster we now have up to 49 values for the turn-around radius.
As a measure of central value and error-bars we use the $50$, $16.8$ and $83.2$ percentiles. Since the distribution of the turn-around radii 
is unknown, 
we also compare with error-bars estimated by fitting a Gauss (as well as an inverse-gamma distribution)
to the distributions of turn-around radii, and we find no
statistically significant change in any conclusions.

\begin{figure}[tbp]
\centering 
\includegraphics[width=.80\textwidth,origin=c,angle=0]{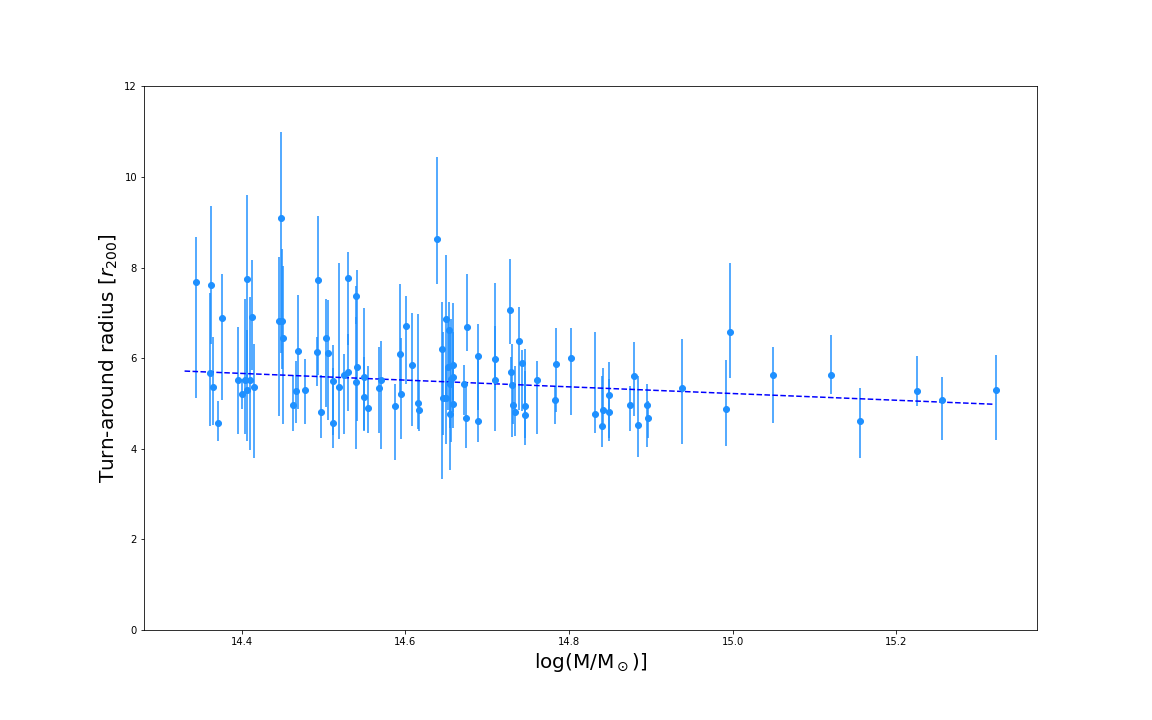}
\caption{\label{fig:rtavsmass} The measured turn-around radius as a function of 
virial mass for
100 massive galaxy clusters from a $\Lambda$CDM 
simulation.  
The data can be approximated with a line of the shape in equation~(\ref{eq:rtaM})
using 
$r_{15} = 5.2 \pm 0.1$ and $\alpha_r = -0.74 \pm 0.4$.}
\end{figure}

\begin{figure}[tbp]
\centering 
\includegraphics[width=.80\textwidth,origin=c,angle=0]{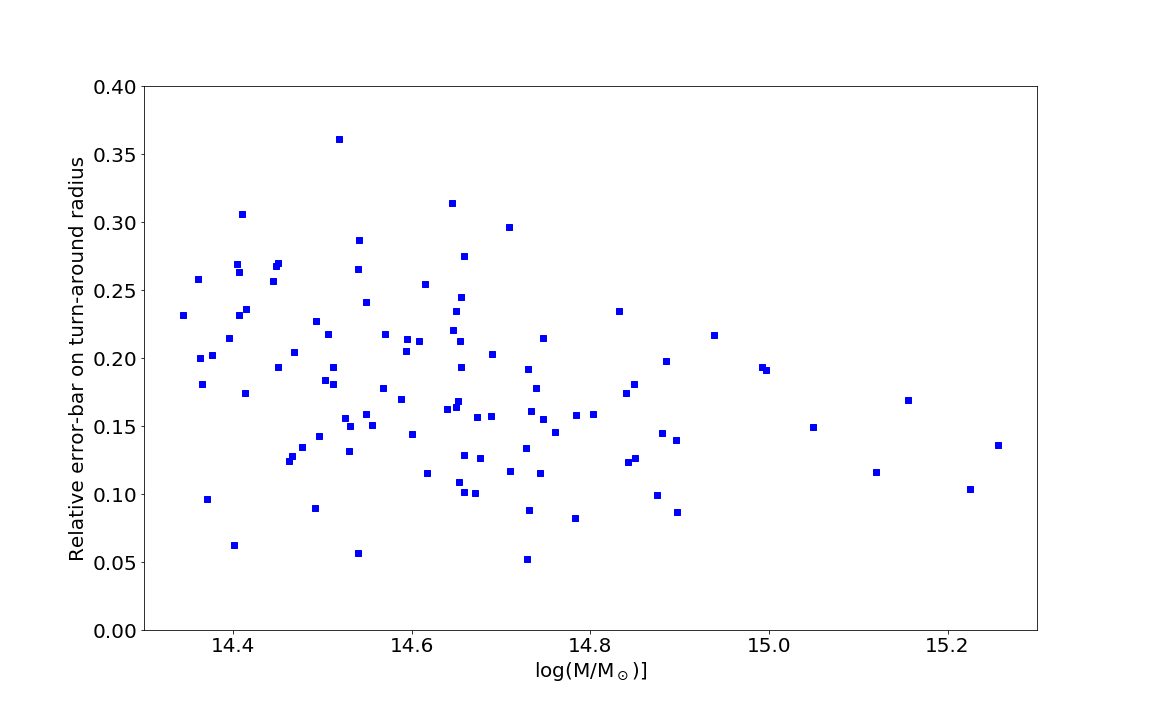}
\caption{\label{fig:errorbars} The relative error-bars on
the turn-around radius as a function of mass for a 
$\Lambda$CDM simulation. The figure shows the symmetrized
error-bars  divided by the central value of
$r_{\rm ta}$.}
\end{figure}

We select 100 massive clusters in the mass-range $10^{14.2} - 10^{15.4} M_\odot$
in a given numerical cosmological
simulation. Considering first a standard $\Lambda$CDM cosmological simulation,
we plot the central turn-around radius with error-bars
as a function of virial mass in figure \ref{fig:rtavsmass}. From this
figure we see two things, first of all that the error-bars are rather
significant for any given cluster, typically of the order $20\%$. For equilibrated
and fairly isolated clusters this may be as low as $10\%$, and for less
equilibrated clusters as high as $40\%$.
And
second, that there are large variations from cluster to cluster, even
for similar mass clusters. The relative error-bars
are shown in figure~\ref{fig:errorbars} for 100 clusters from
a $\Lambda$CDM simulation.

The trend of the mass-dependence of the turn-around radius
can be approximated with a line of the shape
\begin{equation}
    r_{\rm ta} = r_{15} + \alpha_{\rm r} \times {\rm log} \left( \frac{M}{10^{15} M_\odot} \right) \, .
    \label{eq:rtaM}
\end{equation}

{ Today less than 10 Zeldovich pancakes have been
  identified, and yet we are considering using 100 as a future goal.
  In principle one could generalize the statistical analysis in the
  present paper by varying the number of clusters. We leave that for a
  future analysis.}

\section{Various cosmologies}

In order to quantify the capability of the turn-around radius to constrain non-standard cosmologies, we
shall consider
three
different classes of scalar field theories embedded
in the action
\begin{equation}
 S = \int d^4x\sqrt{-g} \left[ \frac{R}{2\kappa^2} +K(\phi,X) \right] + S_c[\psi_c;A_c^2(\phi)g_{\mu\nu}] + S_b[\psi_c;A_b^2(\phi)g_{\mu\nu}] \,, \label{eq:genaction}
\end{equation}
where $K(\phi,X)$ is a free function of the scalar field $\phi$ and $X\equiv-\frac{1}{2}\partial^{\mu}\phi\partial_{\mu}\phi$.
$A_c^2(\phi)$ and $A_b^2(\phi)$ describe couplings of the cold dark matter and baryonic matter fields, $\psi_c$ and $\psi_b$, to the metric field $g_{\mu\nu}$ defining the Ricci scalar $R$.
Furthermore, $\kappa^2 \equiv 8\pi G$ with bare gravitational coupling $G$.

In the following, we shall briefly introduce the three specific models of interest:
quintes\-sen\-ce (Sec.~\ref{sec:quintessence}) and $k$-essence (Sec.~\ref{sec:kessence}) dark energy and a scalar field interaction between cold dark matter particles (Sec.~\ref{sec:fRgravity}).
%
In Sec.~\ref{sec:sims}, we will then describe the
numerical implementations and simulations.

\subsection{Quintessence} \label{sec:quintessence}

Quintessence~\cite{Wetterich:1987fm,Ratra:1987rm} theories are the archetypal dark energy models and are described by Eq.~\eqref{eq:genaction} with the choice of $K(\phi,X)=X-V(\phi)$ with a canonical kinetic contribution, linear in $X$, and the scalar field potential $V(\phi)$.
The models are minimally coupled to the matter sector with $A_b=A_c=1$.
The freedom in choosing $V(\phi)$ can be represented by the freedom in choosing the time dependent dark energy equation of state $-1<w(t)\leq1$, where
\begin{equation}
 w = \frac{\dot{\phi}^2-2V}{\dot{\phi}^2+2V}
\end{equation}
and dots indicate derivatives with respect to cosmological time $t$.

\subsection{{\em k}-essence} \label{sec:kessence}

The $k$-essence~\cite{ArmendarizPicon:1999rj} models describe a class of more exotic dark energy models with noncanonical kinetic contributions in $K(\phi,X)$, nonlinear in $X$, that are minimally coupled to the matter fields ($A_b=A_c=1$).
The freedom in the choice of function $K(\phi,X)$ introduces an additional freedom over the quintessence models with the squared nonluminal sound speed of scalar field fluctuations
\begin{equation}
 c_s^2 = \frac{K_X}{2X \, K_{XX} + K_X}
\end{equation}
in addition to
\begin{equation}
 w = \frac{K}{2X \, K_X - K} \,,
\end{equation}
where subscripts of $X$ denote derivatives with respect to $X$.

\subsection{Scalar dark sector interactions} \label{sec:fRgravity}


As a third example, we study the interaction of cold dark matter with the scalar field $\phi$, specified by the choices $K(\phi,X)=X-V(\phi)$, a minimal coupling to baryons $A_b=1$, and nonminimally coupled dark matter particles $A_c=A(\phi)$.
We choose an interaction
\begin{equation}
 A^2(\phi) = 1 + \sqrt{\frac{2}{3}}\kappa\phi \label{eq:dsi}
\end{equation}
and a potential
of the form
\begin{equation}
 V(\phi)  =  V_0 + V_1 \sqrt{\kappa \phi} \label{eq:dsipotential}
\end{equation}
with
\begin{equation}
 V_0 = \frac{\Lambda}{\kappa^2} \,, \quad V_1 = -\frac{\bar{R}_0}{\kappa^2} \left( \sqrt{\frac{2}{3}}  {\chi}_0 \right)^{1/2} \,, \label{eq:dsipotentials}
\end{equation}
where $\bar{R}_0$ is the Ricci scalar evaluated at the current cosmological background and ${\chi}_0$ is the model parameter with $\chi\equiv\sqrt{2/3}\kappa\phi$.
For ${\chi}_0\ll1$ the background matches that of $\Lambda$CDM, and for our analysis we shall adopt the parameter value ${\chi}_0=10^{-4}$.

Note that we have chosen the model such that the dark sector interaction reduces to the Hu-Sawicki ($n=1$) $f(R)$ gravity model~\cite{Hu:2007nk} in the absence of baryons ($S_b=0$).
In this case, ${\chi}_0 = - \left.(df/dR)\right|_{R=\bar{R}(z=0)} \equiv - \left.f_R\right|_{R=\bar{R}(z=0)} \equiv - f_{R0}$.
The correspondence follows from applying the conformal transformation of the metric $\tilde{g}_{\mu\nu}=A^2(\phi)g_{\mu\nu}$ in the limit of $\chi\ll1$ ($\phi\ll1$).
The resulting coupling to the Ricci scalar can be cast as $f_R R$ such that the Lagrangian density of the gravitational sector becomes $\mathcal{L}_g=R+f(R)$ with $f(R) = -2\Lambda - f_{R0}\bar{R}_0^2/R$.
For simplicity, we will assume here that all matter is in the form of cold dark matter such that the models become equivalent.
Stringent Solar-System constraints on $f(R)$ gravity, relying on a baryonic coupling, however, no longer apply.
Cosmological constraints~\cite{Lombriser:2014dua} such as from the abundance of clusters that are largely independent of the baryonic coupling require $\chi_0 \lesssim (10^{-5}-10^{-4})$~\cite{Schmidt:2009,Lombriser:2010mp,Cataneo:2016iav}.

The choice of interaction~(\ref{eq:dsi}) and potential~(\ref{eq:dsipotential}) induces a chameleon screening mechanism for deep gravitational potentials $\left|\Psi_{\rm N}\right| \gg 3\left|\delta\chi\right|/2 \equiv 3\left|\chi-{\chi}_0\right|/2$.
The potential wells for the structures considered in this work are, however, significantly weaker (${\chi}_0=10^{-4}$).
Screening effects can therefore be neglected, and we can adopt a linearisation of the interaction.
More specifically, we linearise the quasistatic scalar field equation around the cosmological background such that~\cite{Schmidt:2008tn,Lombriser:2012nn}
\begin{equation}
 \nabla^2 \delta\chi - m^2\delta\chi = \frac{\kappa^2}{3}\delta\rho_m \,,
\end{equation}
where the scalar field mass of the Yukawa interaction is given by
\begin{equation}
 m^2 = \frac{1}{6{\chi}_0} \frac{\bar{R}^3}{\bar{R}_0^2} \,.
\end{equation}
For the Poisson equation 
\begin{equation}
 \nabla^2\Psi_{\rm N} = \frac{\kappa^2}{2} \delta\rho_m + \frac{1}{2}\nabla^2\delta\chi \,,
\end{equation}
this implies an enhanced effective gravitational coupling for the cold dark matter particles constituting $\delta\rho_m$, which in Fourier space is given by
\begin{equation}
 \frac{k^2}{a^2} \Psi_{\rm N} =- \left(1 + \frac{1}{3} \frac{k^2}{k^2 + m^2 a^2} \right) \frac{\kappa^2}{2}\delta\rho_m \,. \label{eq:fRPoisson}
\end{equation}

\section{Numerical simulations} \label{sec:sims}
To investigate the power of using the turn-around radius in distinguishing different cosmologies, 
we run simulations
for $k$-essence dark energy (Sec.~\ref{sec:kessence}) models and a scalar field interaction in the dark sector that reduces to linearised $f(R)$ gravity in the absence of baryons. These can then be compared with a standard $\Lambda$CDM
simulation. The initial conditions for the simulations are set using the linear transfer functions from the linear Boltzmann code CLASS \cite{2011JCAP...07..034B} at high redshift (z=100). All simulations use the same seeds as initial conditions. 
Our $\Lambda$CDM simulations are performed using the gevolution code, which is a relativistic particle-mesh N-body code with a fixed resolution \cite{2016JCAP...07..053A}. 
For the $k$-essence and quintessence simulations we have used the $k$-evolution code, which is a relativistic N-body code based on gevolution, in which the $k$-essence scalar field and Einstein's equations are solved to update the particles' positions and momenta. Detailed tests of this code will be presented in~\citep{kevolution, parametrisation}.

The simulation of the scalar dark sector interaction is performed through an implementation of the modified Poisson equation~\eqref{eq:fRPoisson} in a Newtonian $N$-body code based on gevolution~\cite{Hassani:2019}, which has been validated against the results of Ref.~\cite{Schmidt:2008tn}.

We have thus four different simulations (in addition to $\Lambda$CDM) 
to quantify the effects of a different background, clustering of a $k$-essence scalar field and 
the effect on the clustering coming from the 
scalar dark sector interactions
on  the turn-around radius.
All simulations have a comoving boxsize of $L=300$ Mpc/h, 
and discretize the fields on a grid of linear size N$_\text{grids}$=512, giving a length resolution of $0.58$ Mpc/h. The dark matter phase space is sampled by N$_\text{pcl} = 512^3$ particles, corresponding to a mass resolution of $1.74 \times10^{10}$ \(\textup{M}_\odot\)/h. The detailed parameters of each simulation are shown
in table \ref{table:1}. 

\begin{table}[htb]
\centering
\begin{tabular}{|c| c c c c c|} 
 \hline 
  &   $\Lambda$CDM  &  SDSI & quintessence & quintessence & $k$-essence \\ [0.5ex] 
 \hline\hline
  $k_{pivot}$ [$\frac{1}{\text{Mpc}}$]  &   0.05  &  0.05  &   0.05 & 0.05 & 0.05 \\ 
  $A_s$ & $2.215 \times 10^{-9}$ &  $2.215 \times 10^{-9}$ & $  2.215 \times 10^{-9}$ & $2.215 \times 10^{-9}$ & $2.215 \times 10^{-9}$ \\
 $\Omega_b h^2$ &   0.022032 &  0.022032 &  0.022032 & 0.022032 & 0.022032 \\
 $\Omega_{cdm} h^2$ &   0.12038 &  0.12038 &  0.12038 & 0.12038 & 0.12038 \\
 $T_{cmb}[K]$ &  2.7255 &   2.7255 &   2.7255 &  2.7255 &  2.7255 \\
 $N_{ur}$ &   3.046 &  3.046 &  3.046 & 3.046 & 3.046 \\
 $c_s^2 $ &  -- &  -- &  \cellcolor{pink}  1 &  \cellcolor{pink}  1 &  \cellcolor{pink}  $-10^{-7}$ \\
  $\Omega_{\Lambda} $ &  \cellcolor{pink}   0.687862 &  \cellcolor{pink}   0.687862 &  -- & -- & -- \\
 $\Omega_{de} $ & -- & -- &  \cellcolor{pink}  0.687862 &     \cellcolor{pink}  0.687862 &     \cellcolor{pink}  0.687862 \\
 $w_{de}$ &   -- &  -- &   \cellcolor{pink}  -0.9 &  \cellcolor{pink}  -0.8 &  \cellcolor{pink}  -0.9 \\
  $\chi_0$ &  --  &   \cellcolor{pink}  $10^{-4}$ &  -- & -- & -- \\
  Initial redshift  &  100  &  100 &  100 & 100 & 100 \\
 [1ex] 
 \hline
\end{tabular}
\caption{The table shows the full information of the simulations, the red color shows where the parameters are changed in different simulations.
In the absence of baryons the scalar dark sector interaction (SDSI) model matches a linearised Hu-Sawicki ($n=1$) f(R) 
gravity model with $\chi_0=|f_{R0}|$.
Note that the imaginary sound speed for $k$-essence is simply chosen to maximise phenomenological modifications in the simulations.
}
\label{table:1}
\end{table}

\subsection{Results}

\begin{figure}[tbp]
\centering 
\includegraphics[width=.80\textwidth,origin=c,angle=0]{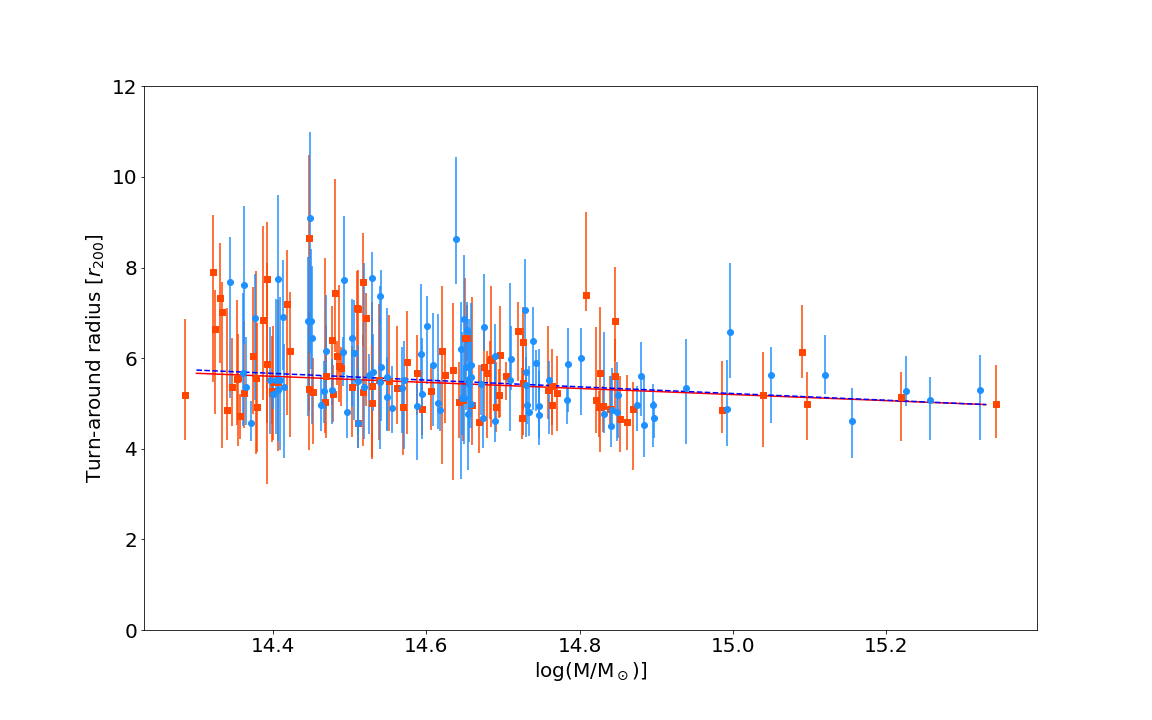}
\caption{\label{fig:kess09} $\Lambda$CDM v.s. $k$-essence with $w=-0.9$
and
$c_s^2=1$ (corresponding to a quintessence model). 
The turn-around radius for the quintessence model
with $w=-0.9$ (red symbols,  red solid line) 
is seen to have essentially the same 
dependence on mass as $\Lambda$CDM (blue symbols, blue dashed line).
These two models cannot be distinguished when measuring the
turn-around radius for 100 galaxy clusters.}
\end{figure}

\begin{figure}[tbp]
\centering 
\includegraphics[width=.80\textwidth,origin=c,angle=0]{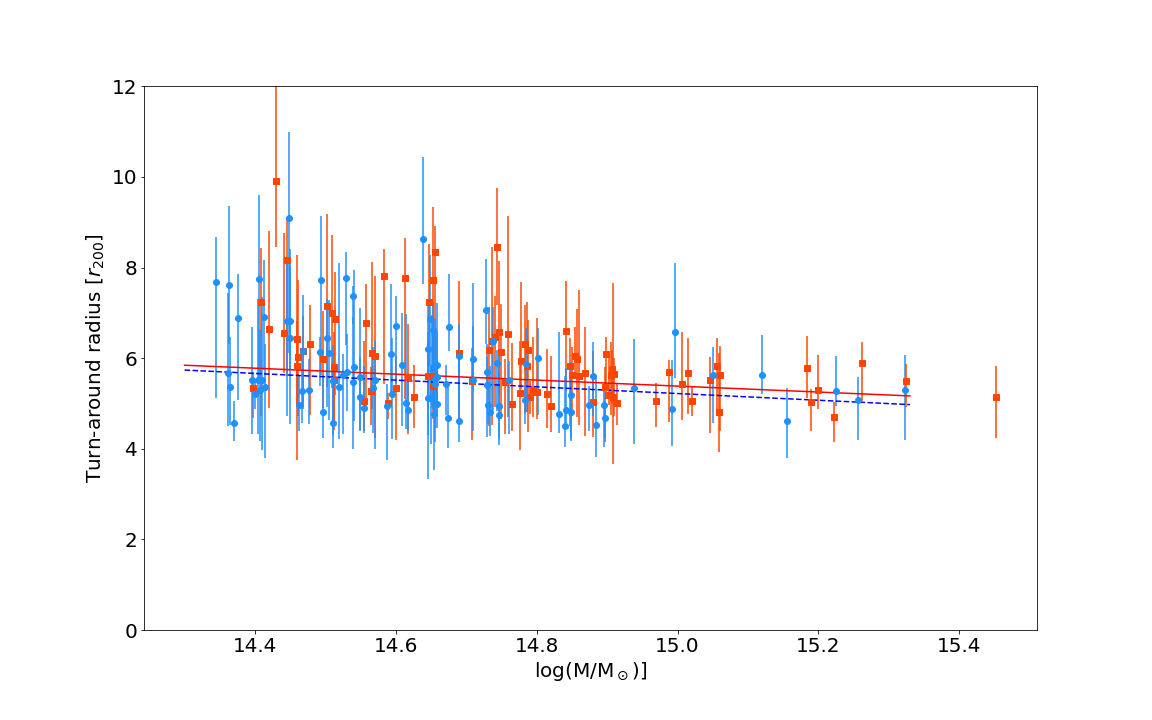}
\caption{\label{fig:fr4} $\Lambda$CDM v.s. scalar dark sector interaction (SDSI). 
{The turn-around radius for the SDSI model (red symbols, red solid
line) is seen to have moved to slightly higher turn-around radius for
the same mass, when compared to $\Lambda$CDM (blue symbols, blue
dashed line).}
The SDSI mass-dependence of the virial mass may be approximated
with a straight line of the form in equation~(\ref{eq:rtaM}), using
$r_{15} =5.4 \pm 0.08$ and $\alpha_{\rm r} = -0.7 \pm 0.3$.
Measuring the turnaround radius
for approximately 100 clusters will allow one to distinguish the two
cosmologies.}
\end{figure}

We compare two of the non-trivial cosmologies with 
a $\Lambda$CDM simulation in figures \ref{fig:kess09} 
and \ref{fig:fr4}.
The first impression is that it will be very difficult
to distinguish between different cosmologies, because of the
large error-bars for each cluster, and the large
cluster to cluster variation. To that
end a careful statistical analysis is needed.

{
One might wonder if it would be advantageous to consider galaxy
groups (or galaxies \citep{2014JCAP...09..020P}) to perform this analysis. That is, however,
still not possible, because the turn-around radius is still only
measurable using the Zeldovich pancake method, which only works near
galaxy clusters.  The reason is, that the Zeldovich pancake method
relies on the gravitational perturbation exerted by the cluster on the
nearby galaxy flow.}


In order to address the problem of cosmic variance, we ran
3 extra numerical simulations of standard $\Lambda$CDM 
universes, with different random seeds for the initial conditions.
By comparing the analysis of each of these universes against the
prediction from our first simulation we can estimate the 
magnitude of variance between different representations of
the same cosmology. To that end we fit the first $\Lambda$CDM 
simulation result by a fit of the shape given in 
eq.~(\ref{eq:rtaM}), $r_{\rm ta}^{\rm fit}$. This is then used in a
chi-squared comparison where we use a sum over the clusters

\begin{equation}
\chi^2 = \sum_{\rm i} \frac{\left( r_{\rm ta}^i - r_{\rm ta}^{\rm fit} \right)^2 }{\sigma_i ^2} \, .
\end{equation}
Here we use symmetrized error-bars for $\sigma_i$
(average of upper and lower error-bars).
Our first simulation has $\chi^2$ per degree of freedom of $0.8$, 
indicating that the error-bars are reasonable.

Comparing with each of the other $\Lambda$CDM simulations leads to
$\Delta \chi^2$ between 1 and 6. This implies that when any given
cosmology is contrasted with the  $\Lambda$CDM, then any $\Delta \chi^2$
less than approximately 6 will not allow us to distinguish the two.

When we perform the same statistical estimator for the simulations 
of quintessence or $k$-essence (see the cosmological parameters in 
table~\ref{table:1})
we get $\Delta \chi^2$ less than 6 for all the cosmologies. This implies that none of the cosmologies
will be distinguishable from $\Lambda$CDM.


The same statistical estimator for the simulations of
SDSI cosmology gives $\Delta \chi^2=9.2$, and thus indicates that it will (as the only
one amongst the ones considered here) be distinguishable from $\Lambda$CDM.
Formally one might think that  $\Delta \chi^2=9.2$ implies that for the two free parameters
used in the fit, the two cosmologies are distinguishable at $99\%$ CL, however,
that estimate does not include the cosmic variance, and the real CL is
therefore smaller.  A careful analysis of the statistics would
require a larger number of simulations and is beyond the scope
of this paper.


The entire analysis presented above has been based on observations
at redshift zero. It is clear that different cosmologies have
different evolution and structure formation, and it is therefore
likely that an analysis including the redshift dependence
will lead to somewhat stronger constraints than what we have 
obtained here.

\section{Conclusions}
The environments of galaxy clusters are complex distributions of
sub-structures, filaments, sheets and voids. This implies that the
turn-around radius, where the radial velocity of galaxies is zero,
varies when different directions in space are considered. 
This implies that one must include a systematic error-bar when
measuring the turn-around radius  in the future. 
We use $\Lambda$CDM numerical
simulations to quantify the magnitude of this error-bar, and
we find that it is about $20\%$ of the measured turn-around radius
for typical clusters, going down to about $10\%$ for the most
equilibrated and spherical structures. Furthermore, we show
that one must carefully avoid measuring the turn-around
radius along directions with large sub-structures.

We  use a range of non-trivial cosmological simulations to gauge
to which extent the inclusion of this realistic error-bar
allows one to measure a departure from a $\Lambda$CDM universe,
and we find that it becomes possible only for the most
extreme cosmologies we considered, such as scalar dark sector 
interaction with fairly large interactions.

\acknowledgments

It is a pleasure to thank Mona Jalilvand and Julian Adamek for support with the
numerical simulations.
SHH is grateful to  the Swiss National Science Foundation and
Carlsberg Foundation for supporting his visit to 
University of Geneva.
This project is partially funded by the 
Danish council for independent research, DFF 6108-00470.
L.L.~acknowledges support by a Swiss National Science
Foundation Professorship grant (No.~170547).
FH and MK acknowledge funding by the Swiss National Science Foundation. This work was supported by a grant from the Swiss National Supercomputing Centre (CSCS) under project ID s710.



\end{document}